\documentclass{ws-procs9x6}
\usepackage{pstricks}

\newcommand{\newangle}{\sphericalangle}
\begin{document}

\title{First measurement of interference fragmentation on a transversely polarized hydrogen target}

\author{P.B. van der Nat \\ (on behalf of the HERMES collaboration)}

\address{Nationaal Instituut voor Kernfysica en Hoge-Energiefysica (NIKHEF), \\
P.O. Box 41882, 1009 DB Amsterdam, The Netherlands \\
E-mail: natp@nikhef.nl}

\maketitle

\abstracts{
The HERMES experiment has measured for the first time single target-spin asymmetries in semi-inclusive two-pion production using a
transversely polarized hydrogen target. These asymmetries are related to the product of two unknowns, the transversity
distribution function and the interference fragmentation function.
In the invariant mass range 0.51 GeV $< M_{\pi^+\pi^-} < 0.97 $ GeV the
measured asymmetry deviates significantly from zero, indicating that two-pion semi-inclusive deep-inelastic scattering can be
used to probe transversity.}

\section{Introduction}
An important missing piece in our understanding of the spin structure
of the nucleon is the transversity distribution $h_1(x)$. It is the only one of
the three leading-twist quark distribution functions, $f_1(x)$, $g_1(x)$
and $h_1(x)$, that so-far
remains unmeasured. The function $h_1(x)$ describes the distribution of transversely
polarized quarks in a transversely polarized nucleon.
It is quite difficult to measure $h_1(x)$, since it is a chiral-odd function, which can
only be probed in combination with another chiral-odd function.
This can be done in semi-inclusive DIS, where the second chiral-odd object is
a fragmentation function, describing the fragmentation of the struck
quark into one or more final-state hadrons.

HERMES is one of the pioneering experiments on this subject. The structure function
$h_1(x)$ is probed by measuring various single-spin asymmetries. First, a
longitudinally polarized target \cite{HERMES_longpol} was used and more recently a transversely
polarized target was used \cite{HERMES_transpol}. In these experiments, single spin asymmetries (SSA's)
were only measured for \emph{single-hadron} semi-inclusive DIS (SIDIS). However, already in
1993 Collins et al. \cite{Collins1993} and in 1998 Jaffe et
al. \cite{Jaffe} suggested to study transversity in
two-hadron SIDIS. Although this comes at the expense of a larger
statistical uncertainty, there is a good reason for looking at SSA's in
two-hadron SIDIS: the measured SSA's relate directly to the product of
$h_1(x)$ and the fragmentation function, whereas in single-hadron SIDIS
this product is convoluted with the transverse momentum of the hadron. Also measuring SSA's in
two-hadron SIDIS provides an independent method of measuring $h_1(x)$, since it involves a different fragmentation
function as compared to single-hadron SIDIS.

In order to finally extract the structure function $h_1(x)$, one needs to know the value of the involved
fragmentation function. Although this function is also still unknown, it can be cleanly measured in
$e^+e^-$ experiments, such as Belle and Babar.


\begin{figure}[t!]
\begin{center}
\includegraphics[width=8cm]{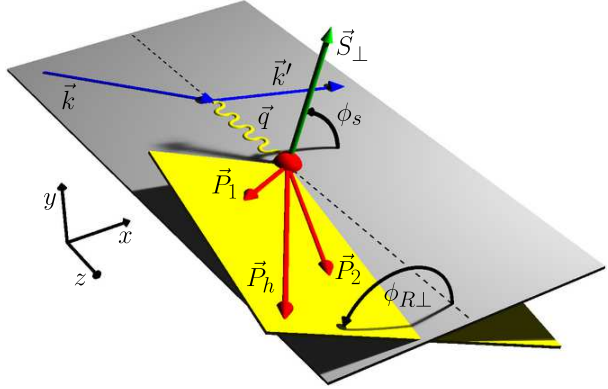}
\end{center}
\caption{Kinematic planes, where $\phi_{R \perp}$ is the angle between the
plane spanned by the incident ($\vec{k}$) and scattered lepton ($\vec{k}'$) and the plane spanned by the two
detected pions $\vec{P}_1$ ($\pi^+$) and $\vec{P}_2$ ($\pi^-$) with $\vec{P}_h \equiv \vec{P}_1 + \vec{P}_2$.}\label{fig:azimuthal_angle}
\end{figure}

\section{Single Spin Asymmetry}

\begin{figure}[t!]
\begin{center}
\includegraphics[width=6cm]{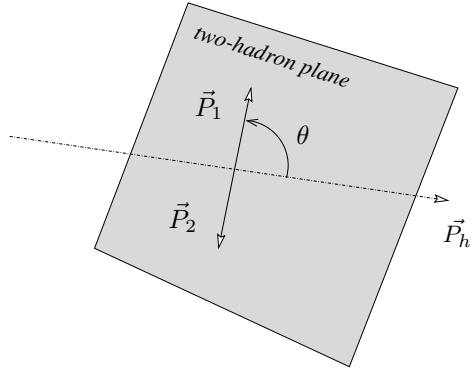}
\rput(-0.15,1.8){\small $\vec{P_h}$}
\rput(-2.2,3.1){$\theta$}
\rput(-3.45,3.5){$\vec{P}_1$}
\rput(-3.8,2  ){$\vec{P}_2$}
\end{center}
\caption{Description of the polar angle $\theta$, in the center-of-mass frame of the
two pions. The vector $\vec{P_h}$ is evaluated in the hadronic center-of-mass system.}\label{fig:polar_angle}
\end{figure}

The transversity distribution can be accessed experimentally by measuring the single target-spin asymmetry,
defined as:
\begin{eqnarray}
A_{UT}(\phi_{R \perp},\phi_S,\theta) & = & \frac{1}{|S_T|} \frac{N^{\uparrow}(\phi_{R \perp},\phi_S,\theta)/N^{\uparrow}_\mathrm{DIS} -
N^{\downarrow}(\phi_{R \perp},\phi_S,\theta)/N^{\downarrow}_\mathrm{DIS}}{N^{\uparrow}(\phi_{R \perp},\phi_S,\theta)/N^{\uparrow}_\mathrm{DIS}
 + N^{\downarrow}(\phi_{R
 \perp},\phi_S,\theta)/N^{\downarrow}_\mathrm{DIS}} \nonumber \\
 & = &  \frac{\sigma_{UT}}{\sigma_{UU}}\mathrm{,}\label{eq:asymmetry}
\end{eqnarray}
where $UT$ refers to Unpolarized beam and Transversely polarized target.
The asymmetry is evaluated as a function
of the angles $\phi_{R \perp}$, $\phi_S$ and $\theta$ which are defined
in Fig. \ref{fig:azimuthal_angle} \footnote{The angle definitions
are consistent with the ``Trento Conventions''
\cite{Trento_conventions}.} and \ref{fig:polar_angle}.
Explicitly:
\begin{equation}
\phi_{R \perp} = \frac{\vec{q} \times
 \vec{k}\cdot\vec{R_T}}{|\vec{q} \times \vec{k}\cdot\vec{R_T}|}
 \cos^{-1} {\frac{
\vec{q} \times \vec{k}\cdot\vec{q}\times\vec{R_T}}{|\vec{q} \times \vec{k}||\vec{q}\times\vec{R_
T}|}}
\end{equation}
and
\begin{equation}
\phi_S = \frac{\vec{q}\times\vec{k}\cdot\vec{S}_\perp}
                   {|\vec{q}\times\vec{k}\cdot\vec{S}_\perp|}
\cos^{-1} {\frac{\vec{q}\times\vec{k}\cdot\vec{q}\times\vec{S}_\perp}
                {|\vec{q}\times\vec{k}||\vec{q}\times\vec{S}_\perp|}}\mathrm{.}
\end{equation}

where $R_T$ is the component of $R$ ($\vec{R} \equiv (\vec{P_1} -
\vec{P_2})/2$) perpendicular to $P_h$ ( $\vec{P}_h \equiv \vec{P}_1 +
\vec{P}_2$), i.e. $\vec{R_T} = R - (R\cdot \hat{P_h})\hat{P_h}$.

The azimuthal angle $\phi_S$ represents the spin direction of the target ``$\uparrow$'' state and $N^{\uparrow (\downarrow)}(\phi_{R \perp},\phi_S,\theta)$
is the number of semi-inclusive $\pi^+\pi^-$-pairs in the target $\uparrow$($\downarrow$) spin state.
These numbers are normalized to the corresponding number of DIS events, $N_{\mathrm{DIS}}^{\uparrow}$ and $N_{DIS}^{\downarrow}$,
respectively. The quantity $|S_T|$
indicates the average target polarization.
The asymmetry is equal to the ratio of $\sigma_{UT}$ and $\sigma_{UU}$, which are the polarized and unpolarized cross sections, respectively.
According to Bacchetta et al. \cite{Alessandro_PRD67} $\sigma_{UT}$ can be written at leading-twist\footnote{See \cite{Alessandro_PRD67} for the sub-leading twist expression.} as:
\begin{eqnarray}
\sigma_{UT} & = & - \sum_q \frac{\alpha^2 e_q^2}{2\pi Q^2  y} (1-y) |
 \vec{S}_{\perp}| \frac{|\vec{R}|}{M_{\pi\pi}} \sin(\phi_{R
 \perp}+\phi_S)\sin\theta h_{1,q}(x) \nonumber \\
  &   & \times  \left[ H_{1,q}^{\newangle,sp}(z,M_{\pi\pi}^2) + \cos\theta H_{1,q}^{\newangle,pp}(z,M_{\pi\pi}^2) \right]
\end{eqnarray}
where $\begin{displaystyle} |\vec{R}| = \frac{1}{2} \sqrt{M_{\pi\pi}^2-4M_{\pi}^2} \end{displaystyle}$ with $M_{\pi\pi}$ the invariant
mass of the pion pair, $M_\pi$ the pion mass and $x$, $y$ and $z$ the standard scaling variables used in semi-inclusive DIS.
The transversity distribution $h_1(x)$ couples to a combination of two-hadron interference fragmentation functions,
$H_1^{\newangle,sp}$ and $H_1^{\newangle, pp}$. These functions describe the interference between different production channels
of the pion pair; $H_1^{\newangle,sp}$ relates to the interference between $s$- and $p$-wave states and
$H_1^{\newangle,pp}$ to the interference between two $p$-wave states.

A two-dimensional fit function of the form
\begin{equation}
f(\phi_{R \perp} + \phi_S, \theta) = p_0 + p_1 \sin({\phi_{R \perp} + \phi_S})\sin\theta \label{eq:fit_function}
\end{equation}
was used to extract from the measured asymmetry the part related to the product $h_1H_1^{\newangle,sp}$, where $p_1 \equiv A_{UT}^{\sin(\phi_{R \perp}+\phi_S)\sin\theta}$.

\section{Results}
The present results are based on data taken in the period from 2002 until 2004 using a transversely
polarized hydrogen target in the HERMES experiment at DESY. The average target polarization, $|S_T|$, was \mbox{75.4 $\pm$ 5.0 \%}.

\begin{figure}[t!]
\begin{center}
  \includegraphics[width=7.5cm]{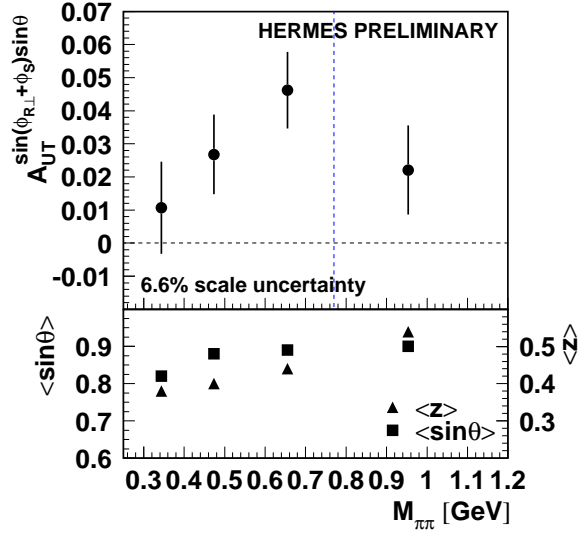}
  \caption{The asymmetry $A_{UT}^{\sin(\phi_{R \perp} +
      \phi_S)\sin\theta}$ versus the invariant mass of the
    $\pi^+\pi^-$-pair (using mass binning, with the bin boundaries at 0.25, 0.40, 0.55, 0.77, 2.0 GeV).}\label{fig:asymmetries1}
\end{center}
\end{figure}
\begin{figure}[t!]
\begin{center}
  \includegraphics[width=6.5cm]{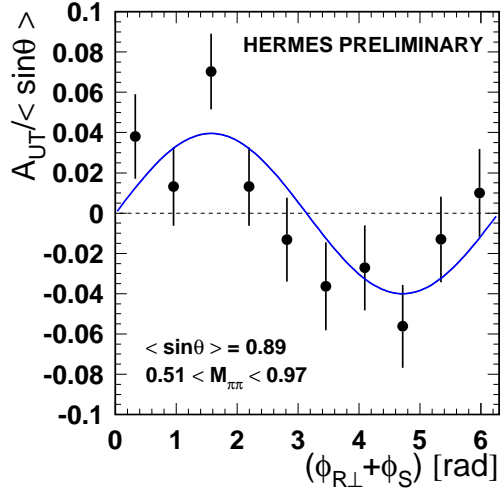}
  \caption{The asymmetry $A_{UT}$ divided by the average $\langle \sin\theta \rangle$ versus the angle combination
$(\phi_{R \perp} + \phi_S)$.}\label{fig:asymmetries2}
\end{center}
\end{figure}
\begin{figure}[t]
\includegraphics[width=5.cm]{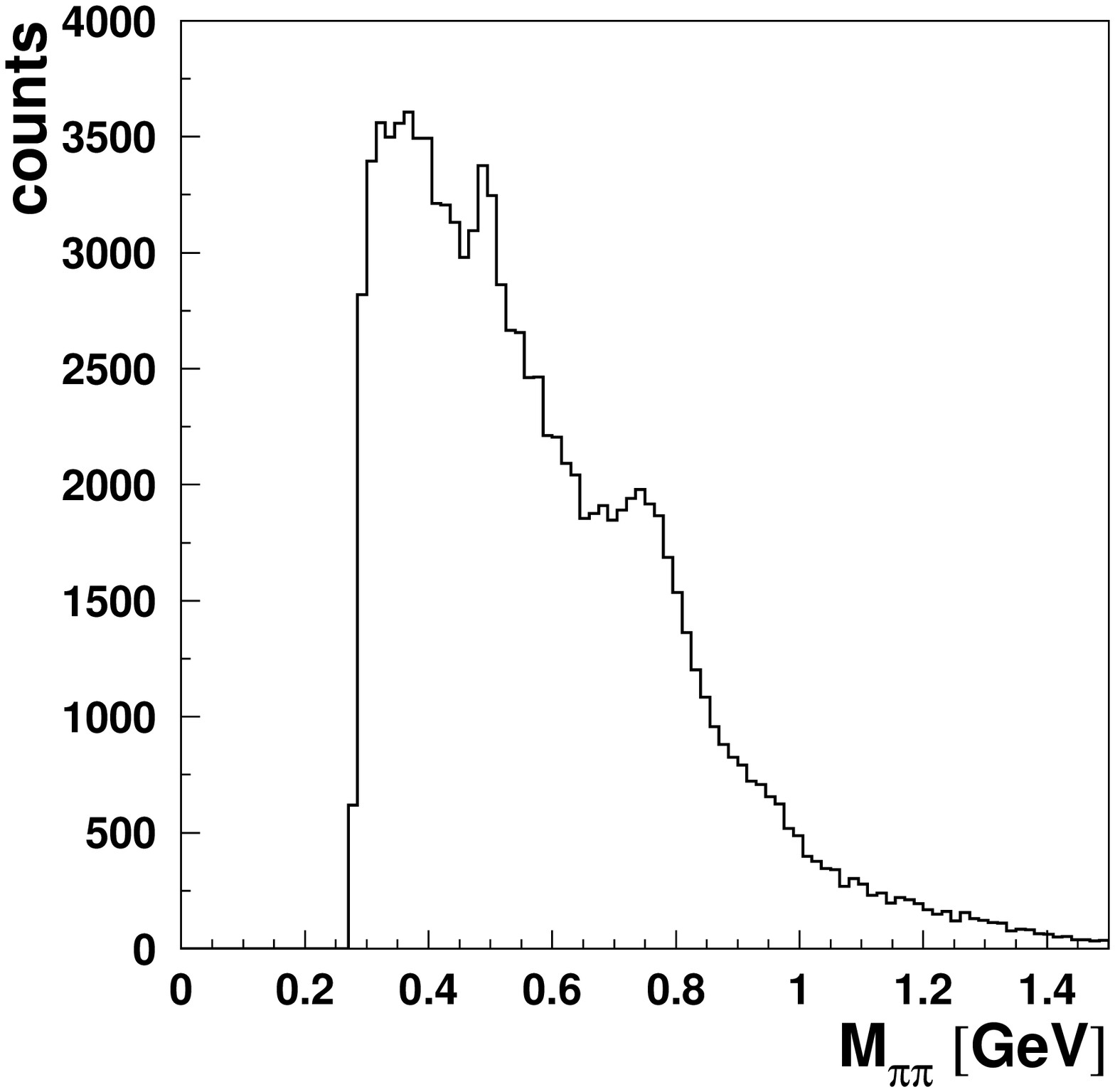}
\rule{.5cm}{.0cm}
\includegraphics[width=5.cm]{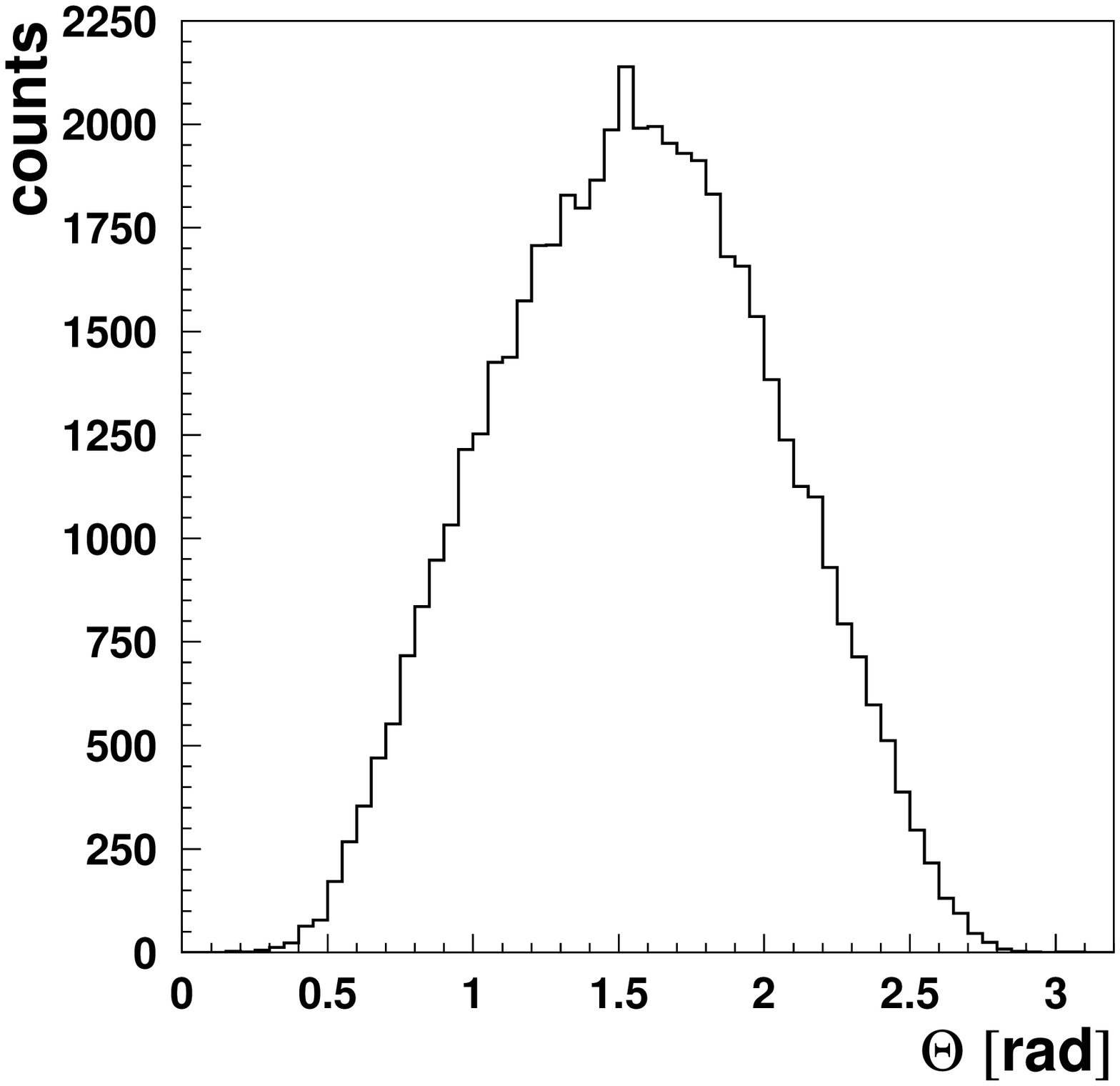}
\caption{The left plot shows the distribution of the invariant mass of the $\pi^+\pi^-$-pairs and the right plot shows the distribution
of the angle $\theta$ (for the invariant mass range 0.51 GeV $< M_{\pi\pi} <$ 0.97 GeV.
\label{fig:distributions}}
\end{figure}

In Fig. \ref{fig:asymmetries1} the data for $A_{UT}^{\sin(\phi_{R \perp} + \phi_S)\sin\theta}$ are shown versus the invariant mass
of the $\pi^+\pi^-$-pair. The asymmetry is clearly positive over the entire invariant mass range and largest in the region of
the $\rho^0$ mass. The corresponding invariant mass distribution is shown in
the left plot of Fig. \ref{fig:distributions}.
Whereas the results on SSA's in two-hadron fragmentation using a
\emph{longitudinally} polarized deuterium target
\cite{vandernat_spin2004} gave a hint of a sign change of the
asymmetry at the $\rho^0$ mass (0.770 GeV) as predicted in \cite{Jaffe},
the new results presented here are clearly inconsistent with such
behavior. However, a good description of the data is given by a refined
version \cite{Radici_theseproc} of a prediction which uses a spectator model
for the fragmentation functions \cite{Radici_PRD65}.

In Fig. \ref{fig:asymmetries2} the raw asymmetry is shown in bins of $\phi_{R \perp}+\phi_S$, integrated over the invariant mass range 0.51 GeV $< M_{\pi\pi} <$ 0.97 GeV.
This plot shows that a clear $\sin(\phi_{R \perp}+\phi_S)$ behavior is present in the data. The plot includes a curve resulting from
fitting the data with $f(\phi_{R \perp} + \phi_S) = p_0 + p_1 \sin({\phi_{R \perp} +
  \phi_S})$, where $p_1 \equiv A_{UT}^{\sin(\phi_{R \perp}+\phi_S)\sin\theta}$
= 0.040 $\pm$ 0.009 (stat) $\pm$ 0.003 (syst).
Due to the peaked shape of the $\theta$-distribution
(right plot in Fig. \ref{fig:distributions}) the asymmetry is mostly evaluated around $\theta = \frac{\pi}{2}$.
Therefore the value of $A_{UT}^{\sin(\phi_{R \perp}+\phi_S)\sin\theta}$ is insensitive
to whether one uses this one-dimensional fit function, integrating over
$\theta$, or a two-dimensional one, like
eq. \ref{eq:fit_function}.

Data taking with a transversely polarized hydrogen target will continue until
November 2005 after which the analysis of the full data sample is expected to lead to a decrease
of the uncertainty on the asymmetry with approximately a factor of
$\sqrt{2}$. Further steps in the analysis include looking at the part of
the asymmetry coupling to $H_1^{\newangle,pp}$ and studying the
$x$ and $z$ dependence of the asymmetries.

\section*{Acknowledgments}
We acknowledge the support of the Dutch Foundation for Fundamenteel Onderzoek der Materie (FOM) and
the European Community-Research Infrastructure Activity under the FP6 ''Structuring the European Research
Area'' program (HadronPhysics, contract number RII3-CT-2004-506078).

\end{document}